\documentclass[showpacs,preprintnumbers,amsmath,amssymb,endfloats*]{revtex4}
\usepackage{graphicx}

\newcommand{\bes}{\begin{eqnarray}}
\newcommand{\ees}{\end{eqnarray}}

\begin{document}

\thispagestyle{empty}
\title{
Experiment and theory in the Casimir effect
}
\author{G.~L.~Klimchitskaya}
\affiliation{North-West Technical University, Millionnaya St. 5, St.Petersburg,
191065, Russia}
 \author{V.~M.~Mostepanenko}
\affiliation{Noncommercial Partnership ``Scientific Instruments'', 
Tverskaya St.{\ }11, Moscow, 103905, Russia 
}

\begin{abstract}
Casimir effect is the attractive force which acts between two plane
parallel, closely spaced, uncharged, metallic plates in vacuum.
This phenomenon was predicted theoretically in 1948 and
reliably investigated experimentally only in recent years.
In fact, the Casimir force is similar to the familiar van der Waals
force in the case of relatively large separations when the relativistic
effects come into play. We review the most important experiments
on measuring the Casimir force by means of torsion pendulum, atomic
force microscope and micromechanical torsional oscillator. Special
attention is paid to the puzzle of the thermal Casimir force, i.e., 
to the apparent violation of the third law of thermodynamics when the
Lifshitz theory of dispersion forces is applied to real metals.
Thereafter we discuss the role of the Casimir force in nanosystems
including the stiction phenomenon, actuators, and interaction of
hydrogen atoms with carbon nanotubes. The applications of the
Casimir effect for constraining predictions of extra-dimensional
unification schemes and other physics beyond the standard model
are also considered.
\end{abstract}
\pacs{12.20.Fv, 12.20.Ds}
\maketitle

\section{Force arising from the vacuum}

In 1948 H.~B.~G.~Casimir \cite{1} made a prediction that two 
large, neutral,
parallel conducting plates separated by a distance $z$ in vacuum
attract each other with the force per unit area
\[
P(z)=\frac{F(z)}{S}=-\frac{\pi^2}{240}\,\frac{\hbar c}{z^4}.
\]
\noindent
Here $\hbar$ is the Planck constant, $c$ is the velocity of light,
and $S$ is the area of the plates.
Figure 1 illustrates configuration giving rise to this force which
was named after Casimir. The Casimir force is a quantum phenomenon
(because it depends on $\hbar$) and also the relativistic one
(because it depends also on $c$). In classical
electrodynamics, there is no net force acting between uncharged
conducting plates. Thus the Casimir force is very unusual. All forces
which we know from both classical and quantum physics depend on some
charges or interaction constants. For example, electric force acts 
between charged bodies and depends on their charges.
The gravitational force depends on the masses (i.e., the gravitational
charges) of interacting bodies. Forces acting between elementary
particles depend also on the constants of weak and strong interactions.
But the above expression for the Casimir force per unit area of the
plates (i.e., for the pressure) does not depend on any interaction
constant. It depends on only the {\it fundamental constants}
$\hbar$ and $c$ and on the separation distance $z$ which is the
geometrical parameter. The magnitudes of the Casimir pressure are
characteristic for macroscopic rather than for microscopic scales.
For example, at a separation $z=1\,\mu$m it holds
$P=1.3\,$mPa which is not a small pressure as one could expect
for a quantum phenomenon.

Physical explanation given by Casimir to his effect is connected
with the concept of quantum vacuum. According to quantum field theory,
``empty'' space is in fact filled with zero-point (or vacuum)
oscillations of all frequencies. These oscillations are sketched in
figure 2, left. The total energy of the vacuum oscillations is equal to
infinity. This, however, does not create a problem. The point is that
in all fields of physics (with exception of gravitation) the energies
are defined up to an additive constant. It is generally believed that
all physical energies are measured from the top of the infinite
vacuum energy in empty space. Because of this, when calculating the
physical energy of some quantum process, one arrives first at an infinite
result and then makes it finite by subtracting the infinite vacuum
energy in empty space.

Using this procedure, Casimir calculated the total energy of vacuum
oscillations between two parallel plates made of ideal metal and spaced 
a distance $z$ apart. This calculation took into account that on metal
surfaces the tangential component of electric field (it is parallel
to the surface) vanishes. This boundary condition selects the vacuum
oscillations shown in figure 2, right. The selected oscillations
have roots on the plate surfaces and their total energy is equal
to infinity. However, when we subtract from this infinity the infinite
energy of vacuum oscillations in empty space, a finite energy which
depends on $z$ is found. The Casimir force is just the negative derivative 
of the above obtained finite energy with respect to separation.

In essence, the Casimir force can be understood in a unified way with 
the well known van der Waals force. The van der Waals force acts 
between two neutral atoms or molecules separated by a distance which
is rather small but much larger than the atomic dimensions. For
nonpolar molecules, which have no intrinsic dipole moments, the
expectation values for their dipole moment operators are zero.
The van der Waals force arises in second order perturbation theory
from the fluctuating dipole-dipole interaction. It is caused by the 
dispersions of
dipole operators, i.e., by quantum fluctuations which create the
instantaneous dipole moments in atoms and nonpolar molecules.
Thus, the van der Waals force is a quantum phenomenon and depends
on $\hbar$. This force acts also between an atom and a macroscopic
body and between two closely spaced macroscopic bodies because each 
atom of one body interacts with each atom of another. Regarding vacuum
oscillations it can be said that there is a fluctuating electromagnetic
field both inside the condensed bodies and also in the gap between them.

If the separation between two atoms, an atom and a macroscopic body
or between two macrobodies is large enough, the retardation of the
electromagnetic fluctuating interaction comes into play due to the 
finiteness of the velocity of light. In this regime, the van der Waals
force depends on both $\hbar$ and $c$ and is in fact the generalization
of the above discussed above Casimir force, arising due to boundary
conditions on ideal metal plates, for the case of real material
bodies. (Notice that for ideal metals which have infinite conductivity 
the above dependence
$P\sim z^{-4}$ is applicable at all separations,
but for real bodies in a nonrelativistic regime of short separations
$P\sim z^{-3}$ holds.) Thus, both the van der Waals and Casimir
forces are caused by quantum fluctuations. Historically, the name
``van der Waals'' was usually connected with the nonrelativistic case
and the name ``Casimir'' (or ``Casimir-Polder'' for atom-atom or
atom-wall interactions \cite{2}) with the relativistic case. The generic
name for both kinds of interactions is ``dispersion forces''
\cite{3}.

In 1956 E.~M.~Lifshitz \cite{4} developed a more general theory of the
van der Waals and Casimir forces acting between two thick parallel
plates (semispaces) with plane boundary surfaces (see figure 3).
In this theory the plate materials are considered as continuous
media described by the frequency-dependent dielectric permittivity
$\varepsilon(\omega)$ and characterized by randomly fluctuating
sources of the electromagnetic field. On the boundary surfaces
electromagnetic fields obey the familiar boundary conditions of
classical electrodynamics, i.e., that the tangential components
of electric field {\boldmath{$E$}} and magnetic induction
{\boldmath{$B$}} are continuous as well as the normal components of
magnetic induction and electric displacement {\boldmath{$D$}}
(only nonmagnetic materials are considered).    The Lifshitz formula
for the dispersion force takes into account not only vacuum
fluctuations but also thermal photons when the material of the plates
is in thermal equilibrium at some nonzero temperature $T$ (in Casimir
formula the ideal metal plates are supposed to be at zero temperature).
The dispersion force is given in the following symbolic form:
\[
F(z,T)=\sum\limits_{l=0}^{\infty}\int_{0}^{\infty}
dk\,R_l(k,i\xi_l,r_{\|},r_{\bot}).
\]
\noindent
Here $R_l$ is some fixed function which depends on the magnitude of
wave vector projection in the plane of plates $k$, the so called
Matsubara frequencies $\xi_l=2\pi k_BTl/\hbar$, where $k_B$ is the
Boltzmann constant, and on the reflection coefficients 
$r_{\|,\bot}=r_{\|,\bot}(i\xi_l,k)$ for two independent polarizations
of electromagnetic field (indices $\|$ and $\bot$ mean that
{\boldmath{$E$}} is parallel and perpendicular to the plane of
incidence, respectively). These reflection coefficients are in fact
the familiar Fresnel's coefficients but calculated on the imaginary
frequency axis. They account for reflections of electromagnetic
oscillations on the boundary surfaces of the plates (see figure 3).
The Lifshitz theory describes both the van der Waals and Casimir
forces (at small and large separations, respectively), and also
the transition region from one type of dispersion force to another.
In the original Lifshitz paper this theory was applied to
dielectrics and the reflection coefficients were expressed in terms
of dielectric permittivities of the plate material
$\varepsilon(i\xi_l)$. In succeeding years the Lifshitz theory has 
found a lot of applications in the investigation of the van 
der Waals and Casimir forces (see, e.g., reviews \cite{5,6,7} and
monographs \cite{3,8,9,10}). The Casimir effect is an interdicsiplinary 
subject and attracts considerable attention in elementary particle
physics, gravitation and cosmology, condensed matter physics and
atomic physics. Below we discuss only a few selected topics in this
wide field of research which are of great promise in both fundamental
physics and nanotechnology but present severe problems in
understanding and interpretation.

\section{How to measure the Casimir force?}

The measurement of the Casimir force is a very complicated problem
because it is very difficult to hold macroscopic bodies at a separation
of 1\,$\mu$m or less (for small test bodies at larger
separations, force becomes smaller than the sensitivity of most
measuring devices). There are several basic challenges that need to be
overcome for the
measurement of the Casimir force. The surface of any real material
is not perfectly shaped. It is covered by roughness, chemical impurities
and dust particles, thus, making surface properties different of bulk
material properties. Next, given that the force has a very strong 
dependence on separation, precise and reproducible measurement of
separation between the two surfaces is required. It is not an easy
matter to measure separations below one micrometer with sufficient
precision. Furthermore, the electrostatic force due to residual
electric charges on the surface and potential differences
should be kept negligible.

The first attempt to measure the Casimir force between metal plates was
undertaken by Sparnaay \cite{11a} using a force balance based on a spring
balance. Measurements indicated the presence of an attractive force,
but its magnitude was found with about 100\% error which prevented a
good comparison with Casimir prediction. In succeeding years many
measurements of the van der Waals and Casimir force between both metal
and dielectric bodies with plane, spherical and cylindrical surfaces
was performed (see review \cite{7}).
Notable improvements have been made in resolution of the above-mentioned
problems. As a result, the experimental error of force measurement was
decreased to about 50\%. The slow progress was caused by outstanding
difficulties in the measurement of small forces and separations
between closely spaced rough surfaces.

The first in the modern stage of Casimir
force measurements, was performed by Lamoreaux in 1997 \cite{11}. 
In this experiment the Casimir force between a gold coated spherical lens 
and flat plate was measured using a balance based on the torsion
pendulum. Several important improvements have permitted to achieve
higher accuracy than in all previous experiments. First, the residual
potential difference between grounded surfaces was compensated with
application of voltage to the lens. The lens was moved towards the plate
by application of voltage to the piezo. The displacement was measured with 
a laser interferometer with an error of 10\,nm. The calibration of the
setup was done by measuring the electrostatic force due to different
applied voltages at large separations where the Casimir force is very
small.

The approximate theoretical expression for the Casimir force acting
between a perfectly shaped plate and a sphere made of ideal metals 
at zero temperature is given by
\[
F(z)=-\frac{\pi^3\hbar cR}{360z^3},
\]
\noindent
where $R=12.5\,$cm is the sphere radius. The geometrical corrections
to this expression are less than $z/R$ \cite{12,13} and, thus, it works
good at separations $z\ll R$ (a condition well satisfied in experiment
\cite{11} with large supply). The experimental data were found to be
in agreement with theory at separations of about 1\,$\mu$m with
a $5-10$\% error. This was a major achievement in comparison with
the previous measurements. However, no evidences of the corrections to
the Casimir force due to finite conductivity (i.e., nonideality) of
metal films, surface roughness and nonzero temperature were reported
(the experiment was performed at laboratory temperature of about
300\,K). Above all the Lamoreaux experiment has attracted much experimental
and theoretical attention to the Casimir effect by applying modern
measurement techniques and opening new opportunities for more precise
tests of the theory.

A radically new approach to the precision measurements of the Casimir force
was proposed by Mohideen and his collaborators \cite{14,15,16,17,18,19,20}.
They have employed the atomic force microscope (a micromechanical device
for the investigation of the surface with high resolution) to perform
a set of most definitive experiments on the Casimir effect.
The schematic diagram of the typical experiment for measuring the Casimir
force using an atomic force microscope is shown in figure 4. The Casimir
force acts between a relatively large polystyrene sphere of about
200\,$\mu$m diameter
and a sapphire disk both coated by metallic layer (Al in the
first experiment and Au in the next). Instead of the usually used sharp tip, 
here a sphere is attached to the cantilever of an atomic force
microscope which flexes under the influence of the force. The flexing
of the cantilever leads to the deflection of the laser beam which is
registered by the photodiodes A and B. The separation distance between
a plate and a sphere can be changed with the help of the piezo. The
calibration of the spring constant of the cantilever and determination
of the absolute separations between a sphere and a plate was performed
by means of application different voltages to the plate while the
sphere was grounded. For this purpose the electric force between the
two bodies was repeatedly measured at separations larger than
2\,$\mu$m, where the Casimir force is negligibly small, and compared with
exact theoretical force-distance relation derived for sphere-plate
configuration in classical electrodynamics. In the most refined 
experiment in \cite{20}, the absolute error of separation measurements
as small as 0.8\,nm was achieved. The roughness peaks are often higher 
than 0.8\,nm. This does not prevent a precise determination
of separations which are measured between zero roughness levels.
The absolute error of force measurements was about
$8.5\,\mbox{pN}=8.5\times 10^{-12}\,$N at 95\% confidence which
results in relative error of 1.75\% at the shortest separation of 62\,nm.

High precision achieved in the measurements of the Casimir force by means
of an atomic force microscope placed more stringent requirements upon
the theoretical computations. To compare more precise experimental results
with theory, it was necessary to take into account the nonideality of
a metal covering the test bodies (i.e., the fact that electromagnetic
oscillations penetrate into the metal up to the ``skin depth''). It was
necessary also to account for the surface roughness (in the experiments 
by Mohideen et al. it was carefully investigated by using the atomic
force microscope with a sharp tip). This was done
in \cite{21} where the Casimir force, acting between a sphere and a disk,
was calculated including of both finite conductivity and roughness
corrections. In figure 5 the measured Casimir force as a function of
plate-sphere separation is shown as open squares. The theoretical
force with corrections due to surface roughness and finite conductivity
is shown by the solid line, and without any corrections by the dashed line.
From figure 5 it is clearly seen that the measurements of the Casimir
force by means of an atomic force microscope are sensitive enough to
demonstrate the role of corrections due to the finite conductivity of a
metal and surface roughness.

The quantification of errors and precision in the Casimir force
measurements by means of an atomic force microscope was performed
in \cite{22}. Special attention was paid to 
the sample-dependent variations
of the optical properties, needed for computations using the
Lifshitz formula, due to the presence of grains. Many other factors
which influence the theoretical result were also carefully analysed.
At the shortest separation of 62\,nm the relative theoretical error of
about 1.7\% was found. With the increase of separation, the theoretical
error decreases. Within the limits of both experimental and theoretical
errors, i.e., on the level of $1-2$\% at the shortest separations
depending on the chosen confidence probability, very good agreement 
between experiment and theory was demonstrated.

The Casimir force discussed above acts perpendicular to the surfaces.
A new interesting physical phenomenon discovered in \cite{18,19} is
the lateral Casimir force. This may arise \cite{19} when the bodies 
are asymmetrically positioned or their properties are anisotropic.
The lateral Casimir force also originates from the modification of
electromagnetic vacuum oscillations by material boundaries \cite{23}.
In \cite{18,19} it was measured between a plate and a sphere both
covered by the aligned sinusoidal corrugations of equal periods but
different amplitudes. The schematic of the experiment is shown in
figure 6. The first sphere is attached to the cantilever of an atomic
force microscope. The second sphere is imprinted with the corrugations 
and interacts with the corrugated plate. The addition of the first
sphere is needed to isolate the laser light from the region between the 
two corrugated surfaces. Unlike figure 4, the corrugated plate is mounted 
vertically in order to suppress the effect of the ordinary (normal)
Casimir force.

Theory predicts that the lateral Casimir force should have a sinusoidal
dependence on the phase difference (lateral displacement) between
corrugations on the sphere and a plate. This was confirmed experimentally
in \cite{18,19}. In figure 7 the average measured lateral Casimir
force is shown by solid squares as a function of lateral displacement.
The solid line is the theoretical prediction. The lateral force
amplitude $3.2\times 10^{-13}\,$N is determined with rather large
error of about 24\% because it is much smaller than the normal force.

Several more experiments on the measurement of the Casimir force
were performed in the last few years. Among them one should
mention the experiment \cite{24} making use of the original Casimir
configuration, i.e., two plane parallel metal plates. It is really
hard to maintain plates parallel at a micrometer separation. This
partly accounts for not too high precision of this experiment
(the original Casimir formula for ideal metals was confirmed in the 
limits of 15\% error). We will return to other experiments on the Casimir
force in the following sections devoted to the puzzle of the thermal
Casimir force and applications to nanotechnology.

\section{Puzzle of the thermal Casimir force}

As was mentioned in section 1, Casimir obtained his famous formula
by assuming that the temperature of the plates is equal to zero. 
In \cite{25} the case when ideal metal plates are at some nonzero
temperature $T$ was considered in the framework of thermal quantum
field theory with appropriate boundary conditions. It was shown
that at short separations (or, equivalently, at low temperatures)
the corrections to the original Casimir result due to nonzero
temperature are negligibly small. 
On the contrary, if separations between plates are several
micrometers (or the temperature is high enough) the temperature
correction becomes large. At room temperature $T=300\,$K and plate
separations larger than 6\,$\mu$m the total Casimir pressure is
already equal to the thermal term
\[
P(z)=-\frac{k_BT}{4\pi z^3}\zeta(3),
\]
\noindent
where $\zeta(x)$ is the Riemann zeta function and 
$\zeta(3)\approx 1.202$. This behaviour is usually called
``the classical limit'' because it is determined by thermal
photons \cite{26}. All these results, obtained for the ideal metals
at $T\neq 0$ from the thermal quantum field theory, follow also
from the Lifshitz formula if one uses what is known as Schwinger's
prescription \cite{27}. According to this prescription, one should 
take limit $\varepsilon\to\infty$ in the Lifshitz formula first
and thenperform summation in $l$. As a consequence, for both 
reflection coefficients at zero frequency it follows 
\[
r_{\|}(0,k)=r_{\bot}(0,k)=1
\]
\noindent
and the results of thermal quantum field theory are reobtained.

In the beginning of 2000, the rapid progress in experiment raised
a question: What is the dependence of the Casimir force on the temperature
for the case of real metal boundaries? At first the answer seemed  
simple. There was an opinion shared by many physicists 
that it can be simply solved by the substitution of the dielectric
permittivity of a metal into the reflection coefficients, whereupon
they are substituted into the Lifshitz formula. This was done
in \cite{28} and in \cite{29,30} using different models for the
dielectric permittivity of a metal and created a puzzle which is
not completely resolved up to the present despite repeated attempts
of several research groups in different countries. The basic facts
of the problem are as follows.

In \cite{28} the Drude model was used where the dielectric permittivity
of a metal depends on the frequency as
$\varepsilon_D(\omega)\sim\omega^{-1}$.
From this it follows
\[
r_{\|}(0,k)=1,\qquad r_{\bot}(0,k)=0
\]
\noindent
in drastic contradiction with the above result obtained for ideal metals.
At this point it is necessary to explain why these zero-frequency
reflection coefficients that are of prime physical interest. The thing
is that at large separations (or, alternatively, at high temperatures)
the zero-frequency term of the Lifshitz formula alone determines the
total magnitude of the Casimir force, the other terms being exponentially
small. Thus, in \cite{28}, instead of the classical limit, as obtained for
ideal metals, it follows
\[
P(z)=-\frac{k_BT}{8\pi z^3}\zeta(3).
\]
\noindent
This is a counter intuitive result because one may use plates made
of metals of higher and higher conductivity (i.e.,
approaching to the case of the ideal metal), but nevertheless the Casimir
pressure were only one half the magnitude found between ideal metals.
Paper \cite{28} (and supporting papers \cite{31,32}) also predict 
relatively large thermal corrections to the Casimir pressure between 
real metals at short separations. These corrections are about 500 
times greater than between ideal metals.

Unlike paper \cite{28}, papers \cite{29,30} have used the plasma model
dielectric function with
$\varepsilon_p(\omega)\sim\omega^{-2}$. This leads to quite different
result for the reflection coefficients at zero frequency
\[
r_{\|}(0,k)=1,\qquad r_{\bot}(0,k)=
\frac{\sqrt{c^2k^2+\omega_p^2}-ck}{\sqrt{c^2k^2+\omega_p^2}+ck},
\]
\noindent
where $\omega_p$ is the plasma frequency. When real metal properties are 
approaching 
the ideal one, it holds $\omega_p\to\infty$, and $r_{\bot}(0,k)$
approaches unity, i.e., to the value valid for ideal metals. Thus,
at large separations papers \cite{29,30} predict the same classical
limit as was obtained for ideal metals from the thermal quantum
field theory which is in accordance with physical intuition.
Furthermore, at short separations papers \cite{29,30} predict only small
thermal corrections to the Casimir pressure in qualitative agreement with
the case of ideal metals.

Although both the above approaches to the thermal Casimir effect between
real metals were in sharp mutual contradiction, over a period of time
there were no decisive theoretical or experimental evidences against one 
or the other. Things have changed after the publication of papers
\cite{33,34} where it was proved that the approach using the Drude
model violates the fundamentals of thermodynamics. It is common
knowledge that the thermal Casimir pressure is the negative derivative
of the free energy per unit area with respect to separation.
The negative derivative of the same free energy with respect to temperature
is the entropy of a fluctuating field per unit area of plates.
As was shown in \cite{33,34}, in the case of perfect metallic crystal 
lattices with no impurities the entropy of a fluctuating field at zero 
temperature, computed using the Drude dielectric function, is not equal
to zero. Thus, the third law of thermodynamics (the Nernst heat theorem)
is violated. On the contrary, if the dielectric function of the plasma
model is used, the entropy at zero temperature is equal to zero in
accordance with the Nernst heat theorem. The behaviour of the entropy
as a function of temperature is shown in figure 8 for the case of
gold plates spaced at a separation $z=1\,\mu$m. Solid line is computed
using the plasma model, and the dashed line, which violates
thermodynamics, using the Drude model.

To avoid the contradiction with thermodynamics, paper \cite{35}
applied the Drude dielectric function to metallic lattices with
impurities possessing some nonzero residual relaxation at zero
temperature (for perfect lattices there is no relaxation at zero
temperature). As a result, \cite{35} arrived at the conclusion that
entropy at zero temperature is equal to zero. This, however, does not
solve the problem of the thermodynamic inconsistency of the approach 
using the Drude model, as perfect crystal lattices have 
a nondegenerate dynamic state of lowest energy and, thus, according to
quantum statistical physics, the entropy at zero temperature must be
equal to zero in this idealized situation. This property, however, is
violated by the approach using the Drude model.

The severity of the problem is connected with the fact that the
Drude model presents the same behaviour of  
$\varepsilon_D(\omega)\sim\omega^{-1}$ at low frequencies, as given by
the Maxwell equations, whereas the behaviour
$\varepsilon_p(\omega)\sim\omega^{-2}$ given by plasma model is
characteristic only for the frequency reqion of infrared optics.
If this is the case, the question arises on the correctness of
extrapolation of the latter behaviour to zero frequency.
In this respect a deeper physical understanding of what is meant
by the ``zero'' frequency is required. According to one approach,
we should extrapolate to zero the actual reflection properties of
plate materials at very low, quasistatic frequencies. According
to the other approach, the zero-frequency limit should not be understood
literally because there are no static fields in the vacuum state and
static-field fluctuations. More likely, it should be understood as
a mathematical limit to zero from the region of characteristic frequencies
giving the major contribution to the Casimir force. The uselessness of 
the Drude model for the Casimir effect then becomes clear because
this model is the self-consistent solution of Maxwell equations with the
real current of conduction electrons \cite{36}. This current is
created through the incidence on the conductor of a real electromagnetic
field. It also involves the electric resistance and heating of a metal,
a phenomenon which cannot be caused by the zero-point, vacuum
oscillations. On the contrary, in the infrared, frequencies are
so high that they cannot cause real current (in this region
electric current is pure imaginary \cite{36}). This is the reason
why the plasma model is well adapted for use in the theory
of the Casimir effect.

Recently the arguments against a literal understanding of the
zero-frequency term in the Lifshitz formula received strong support
from the investigation of the Casimir force between two dielectric plates. 
As was shown in paper
\cite{37}, the Casimir entropy of a fluctuating field between dielectrics 
goes to zero when temperature vanishes if the plate materials are
described by $\varepsilon(\omega)$ with some finite static value
$\varepsilon(0)$. If, however, the dc conductivity of dielectrics
is taken into account, this results in a violation of the Nernst
heat theorem \cite{37}. Note that at nonzero
temperature dielectrics really possess some small but nonzero
conductivity at zero frequency. Then it becomes clear that real
material properties at very low frequencies are in fact irrelevant
to the fluctuating phenomena described by the Lifshitz formula.

Although the plasma model approach is free of contradictions with
thermodynamics, it cannot be considered as a final resolution of
the puzzle of the thermal Casimir force between real metals.
The free electron plasma model does not take relaxation into
account and is not applicable in the frequency regions of the
anomalous and normal skin effects. Thus, strictly speaking, it
should not lead to very exact results at separations between
plates above 2\,$\mu$m at room temperature. Below about
250\,nm it is also not so accurate because it disregards
the interband transitions and other
processes taken into account in the optical data for the complex
index of refraction. A more universal description of the thermal
Casimir interaction between real metals free of contradictions with 
thermodynamics is given in terms of the Leontovich impedance
$Z(\omega)$ \cite{36}. This description was elaborated in papers
\cite{38,39} and is based on the Leontovich boundary condition
imposed on metal surfaces
\[
\mbox{\boldmath$E$}_t=Z(\omega)\left[
\mbox{\boldmath$B$}_t\times\mbox{\boldmath$n$}\right],
\]
\noindent
where {\boldmath$n$} is the unit vector normal to the surface and
directed into a metal.
This permits to obtain an alternative representation of the 
reflection coefficients in the Lifshitz formula without use of the
dielectric permittivity. In the case of the normal and anomalous
skin effect, the impedance approach leads to the same values,
unity, for
the reflection coefficients at zero frequency as was the case for ideal 
metals. In the region of the infrared optics, the impedance reflection
coefficients at zero frequency are
\[
r_{\|}(0,k)=1,\qquad r_{\bot}(0,k)=
\frac{\omega_p-ck}{\omega_p+ck}.
\]
\noindent
In the limit of ideal metals ($\omega_p\to\infty$), 
$r_{\bot}(0,k)=1$ again. Within the separation region from 150 to 250\,nm,
the impedance approach is more accurate than the plasma model approach.
It, however, is also an approximation. The Leontovich boundary
condition is valid with the proviso that $|Z|\ll 1$. Because of this, 
the impedance approach is not applicable at short separations between
the plates below the plasma wavelength. Here,
however, the thermal corrections
are very small and the Lifshitz formula at zero temperature can be
used to calculate the Casimir force. 

Recently, in addition to important theoretical information, there have
been experimental results which help to
resolve the puzzle of the thermal Casimir force.
In Refs.~\cite{40,41} the results of two new
precise experiments on the determination of the
Casimir pressure between metal coated parallel plates by means of
micromechanical torsional oscillator have been reported. In fact, 
in the experimental
setup, shown schematically in figure 9, the Casimir force acts
between a large sphere and a plane plate, similar to the
measurements using an atomic force microscope. Here, however, the
vertical separation $z$ between the sphere and the plate was
varied harmonically with time. Due to the presence of the Casimir
force, the resonant frequency of the oscillator has been changed and 
this change was measured. As is known from theory, the change in the
resonant frequency is proportional to the derivative of the force acting
between a sphere and a plate with respect to separation. This derivative,
in its turn, is approximately proportional to the Casimir pressure $P$
between two parallel plates spaced at a separation $z$. (The error
in the determination of $P$ by using this procedure is no larger than
$z/R$, where $R$ is the sphere radius \cite{7,12,13}.)

Separations between the sphere and the plate were determined by using
the formula
\[
z=z_{\rm meas}-(D_1+D_2)-b\theta.
\]
\noindent
Here $z_{\rm meas}$ was measured interferometrically, $D_1+D_2$ was
found by the application of different voltages and measuring the
electrostatic force, the lever arm $b$ was determined optically,
and the rotation angle $\theta$ was determined by measuring the
difference in capacitance between the right and left electrodes
(see figure 9 for the definition of all quantities).
As a result, the absolute error in separation measurements was
reduced to 0.6\,nm. In a wide separation region from 170 to 300\,nm
the relative error in the Casimir pressure varied between 0.55 and
0.60\% at 95\% confidence probability \cite{41}. This is the
distinctive feature of the experiment under consideration when compared 
to previous measurements where high precision was achieved only at
the shortest separations.

The experimental results \cite{40,41} were compared with the
theoretical computations of the Casimir pressure described above
using the Leontovich impedance and the Drude model. Notice that
these approaches are qualitatively different only at zero Matsubara
frequency. As to the contributions from nonzero frequencies, both approaches
find these using tabulated optical data extrapolated to low frequencies 
and obtain practically the same result. In figure 10, left we plot by dots 
the pressure differences $P^{\rm th}-P^{\rm exp}$ versus separation,
where $P^{\rm th}$ is computed using the impedance approach.
In figure 10, right the differences ${\tilde{P}}^{\rm th}-P^{\rm exp}$
are plotted where ${\tilde{P}}^{\rm th}$ is computed using the Drude
model approach. Solid lines show the absolute errors of the pressure
differences found at 95\% confidence. From figure 10, left it follows
that the impedance approach is consistent with data (the plasma model
approach also turns out to be consistent with data within experimental
and theoretical errors \cite{41}). From figure 10, right it is seen
that the Drude model approach is excluded experimentally within the
separation region from 170 to 700\,nm at 95\% confidence. In the
separation region from 300 to 500\,nm this approach is excluded by
experiment at even higher 99\% confidence \cite{41}. Thus, experimentally
the puzzle of the thermal Casimir force between real metals is resolved
in favour of the approaches with nonzero contributions of both
polarizations of electromagnetic field at zero frequency.
At the same time, the thermal effect predicted by these approaches is
very small at short separations and it is yet to be measured.

\section{Casimir effect and nanosystems}

The first paper anticipating the importance of the Casimir effect in
nanosystems was published 20 years ago \cite{42}. However, only
recently, owing to shrinking device dimensions to nanometers, the
role of the Casimir forces in their performance and fabrication has
been generally recognized. At separations below a few ten nanometers 
the Casimir force dominates over other forces. As a result, movable
components in nanoscale devices fabricated at so short separations
often stick together due to the strong Casimir and van der Waals
attraction. This process is referred to as ``stiction''. It may lead
to the collapse of movable elements to the underlying substrate,
permanent adhesion and other undesirable consequences \cite{43}.
Together with the familiar capillary forces, the stiction leads
to poor yield in the fabrication of micro- and nanomechanical
systems. Thus, it would be of much promise to develop systems
with zero or suppressed Casimir force. This aim can be achieved
if one considers that the Casimir force in layered
structures and in closed volumes can be not only attractive but
also repulsive \cite{5,6,7,8,9}. Then, by using the appropriate
design, one may achieve the equilibrium between the attractive
and repulsive contributions, i.e., effectively obtain a nanosystem
with zero Casimir interaction.

Although stiction is a harmful phenomenon, the Casimir force can
also play a useful role in nanosystems actuating the
nanofabricated silicon plate. This was first demonstrated in paper 
\cite{44} where the gold coated sphere was suspended above one 
side of the heavily doped polysilicon plate which can rotate
around a thin road (a more advanced device of this type shown
in figure 9 was later used in \cite{40,41} for the precision
measurement of the Casimir force). When the sphere was moved
closer to the plate, the Casimir force, acting on the plate, tilted 
it about its central axis towards the sphere. Thus, vacuum
oscillations of the electromagnetic field led to the mechanical
motion of the plate demonstrating the first micromechanical
device driven by the Casimir force.

A similar device was used to demonstrate the influence of the Casimir
force on the oscillatory behaviour of microsystems \cite{45}.
The simple model of the Casimir oscillator is shown in figure 11.
It consists of a movable metallic plate subjected to the restoring
force of a spring obeying the Hooke's law, and the Casimir force
arising from the interaction with a fixed metallic sphere. The
Hooke's force is linear in a shift of the plate $\Delta z$, whereas
the Casimir force is strongly nonlinear. As a result, the potential
energy of this microdevice possesses a local and a global minima
separated by the potential barrier. The Casimir force changes the
resonant frequency of oscillations around the local minimum, and
makes oscillations anharmonic \cite{45}. This could be used in
future micro- and nanoelectromechanical systems.

All precise measurements of the Casimir force discussed above were
performed between metal coated test bodies. However, the most
important materials extensively used in nanotechnology are
semiconductors. The conductivity properties of semiconductors
range from metallic to dielectric and have considerable
opportunity for the control and modulation of the Casimir force.
Measurement of the van der Waals and Casimir forces between
dielectrics has always been a problem because of the localized
electric charges on their surfaces and contact potential
differences \cite{7}. Semiconductors with a relatively high
conductivity have an advantage that their surfaces avoid
accumulation of charges but, at the same time, show a typical
dielectric dependence of the dielectric permittivity on frequency
within a range of frequencies contributing to the Casimir force.

First measurement of the Casimir force between a large gold coated 
sphere and single crystal silicon plate was performed by Mohideen et 
al. \cite{20} by using an atomic force microscope. The experimental
results were compared with the Lifshitz theory at zero temperature.
(At separations from 62 to 120\,nm, where the smallest total
experimental error from 0.87 to 5.3\% is achieved, any approach to 
the calculation of the thermal corrections discussed above leads to 
a negligible effect far below the experimental precision.)
In figure 12 we present the differences 
$F^{\rm th}-{\bar{F}}^{\rm exp}$ between theoretical and mean
experimental Casimir force versus separation. Solid lines show the
95\% confidence intervals. As is seen from this figure, theory is in 
a very good agreement with experiment. Later it was demonstrated that
by changing the conductivity properties of a semiconductor plate,
either by doping or due to irradiation by laser light, it is
possible to change the force-distance relation. This opens new
opportunities for the modulation of the Casimir force in
semiconductor micro- and nanodevices.

One more important application of the van der Waals and Casimir
forces to nanosystems is connected with atom-surface interaction.
This interaction attracted much attention during the past few
years in connection with experiments on quantum reflection of
ultracold atoms and Bose-Einstein condensation \cite{46,47}. In
paper \cite{48} the Lifshitz formula describing atom-surface
interaction was generalized for the case of an atom interacting
with a multi-wall carbon nanotube. These nanotubes are the
nanosystems which can be modelled by a graphite cylindrical shell
containing several concentric hexagonal layers. The study of the
van der Waals and Casimir forces between atoms and carbon
nanostructures has become urgent due to the potential use of
single-wall nanotubes for hydrogen storage (see references
in \cite{48}). It is common knowledge that the storage of hydrogen
is the key problem in the hydrogen energetics which offers an
alternative to petroleum. For this reason, any new hydrogen storage
mechanism would be of much importance.
The van der Waals and Casimir forces acting between hydrogen atoms or
molecules and carbon nanostructures play a deciding role in
absorption phenomena and until recently were practically unexplored.
Calculations show \cite{48} that the location of hydrogen atoms
and molecules inside a multi-wall carbon nanotube is energetically
preferable as compared with the outside. This result is rather 
promising for the problem of hydrogen storage in carbon 
nanostructures and calls for further investigation.

\section{Casimir effect as a test for new physics}

Modern science is aware of four fundamental interactions: gravitational,
electromagnetic, weak and strong. The gravitational interaction is
described by the Einstein's general relativity theory which contains
the Newton's law of gravitation as a particular case. The electromagnetic
interaction is described by the Maxwell's equations of classical 
electrodynamics
and, including quantum effects, by quantum electrodynamics.
The weak interaction is only quantum. It is described together with
electromagnetic phenomena in the framework of the Weinberg-Salam
theory of electroweak interactions. The strong interaction is described
by quantum chromodynamics. There is also the so called standard model
which provides an uniform description of the three interactions:
electromagnetic, weak and strong. Many attempts were undertaken
during the last decades to unify all four interactions. They, however,
did not achieve much success. The test of many theoretical predictions 
would require construction of new accelerators of huge energies,
unattainable in the foreseeable future.

Many extensions of the standard model, including supergravity and string 
theory, make use of an old idea that the true dimensionality
of space-time is larger than four. It is supposed that the additional
spatial dimensions are compactified at some length scale, which is so
small that they do not influence our everyday life and even precise
scientific experiments. For a long time it was generally believed that
the compactification scale is on the order of the Planck length,
i.e., $\sim 10^{-33}\,$cm. The corresponding energy scale is of order
$10^{19}\,$GeV.

The situation changed dramatically with the proposal of unification
models for which the compactification energy may be of order
$1\,\mbox{TeV}=10^{3}\,$GeV \cite{49}. According to these models,
the compactification scale of extra dimensions may be as large as
a fraction of millimeter. Such ``large'' extra dimensions are
consistent with observations if it is presumed that the usual
physical fields of the standard model exist only in the ordinary
four dimensional space-time whereas gravity alone propagates into
the extra spatial dimensions. As a result, the Newtonian gravitational
potential between two point masses $m_1$ and $m_2$ (atoms for instance)
separated by a distance $r$  acquires a Yukawa correction
\[
V(r)=-\frac{Gm_1m_2}{r}\left(1+
\alpha_Ge^{-r/\lambda}\right),
\]
\noindent
where $G$ is the gravitational constant, $\alpha_G$ is a dimensionless
constant characterising the strength of the Yukawa force, and 
$\lambda$ is its range. It is significant that the Yukawa-type
forces are also predicted in ways unrelated to extra-dimensional
physics. For example, the Yukawa potential describes new forces
generated by the exchange of light bosons of mass $\hbar/(\lambda c)$
which arise in many other extensions of the standard model \cite{50,51}.

Interestingly, at short separations the gravitational experiments of
E\"{o}tvos- and Cavendish-type do not impose strong constraints on the
values of $\alpha_G$ and $\lambda$. The best constraints obtained from
the gravitational experiments made to date are shown in figure 13
where the region of ($\lambda,\alpha_G$) plane above the lines 1--7 is
experimentally prohibited and below these lines is permitted
(see paper \cite{52} where the references to all experiments used
in this figure are presented). From figure 13 it is seen, for instance,
that for the interaction range $\lambda=10^{-5}\,$m the Yukawa-type
correction with $\alpha_G=10^{5}$ is not prohibited experimentally.
If such correction would exist, the corresponding force would be 
$10^5$ times greater than the 
usual Newtonian gravitation at a separation $r=10^{-5}\,$m. Thus, 
at such short separations gravity loses its role as the dominant force
acting between non-magnetic electrically neutral bodies. For these
separations, constraints on new forces and extra dimensions should be
extracted from the measurements of the van der Waals and Casimir
forces \cite{7,9}. 

The first constraint on new physics from the Casimir force measurements
between metals was obtained \cite{53} by analysing data of the
torsion pendulum experiment \cite{11}. This constraint is shown
by line 7 in figure 13. For $\lambda<5.6\,\mu$m it becomes stronger
than any constraint obtained from gravitational experiments.
The rigorous approach to the constraining hypothetical forces from
the measurements of the Casimir force was developed in \cite{41,52}.
In the second improved experiment by means of a micromechanical
torsional oscillator \cite{41,52}, the differences between the
theoretical Casimir pressures, calculated using the Leontovich
impedance, and the experimental ones belong to the 95\% confidence
interval with a half-width 
$\Delta^{\!\rm tot}[P^{\rm th}(z)-P^{\rm exp}(z)]$ shown by the
solid lines in figure 10, left. The quantity $\Delta^{\!\rm tot}$ is
in fact a characteristic value of the agreement between experiment
and theory at a chosen confidence. If in addition to the Casimir
pressure there were some hypothetical pressure $P^{\rm hyp}(z)$ due to
extra dimensions or exchange of light elementary particles between the
test bodies, its magnitude should be less or equal to the
half-width of the confidence interval
\[
|P^{\rm hyp}(z)|\leq\Delta^{\!\rm tot}[P^{\rm th}(z)-P^{\rm exp}(z)].
\]

This inequality permits to obtain constraints on new forces. Really,
the hypothetical pressure $P^{\rm hyp}$ between the two plates at
a separation $z$ can be found as a negative integral of the interatomic
Yukawa potentials over the volumes of plates with subsequent
differentiation of the obtained result with respect to separation.
Thus, this pressure depends on $\alpha_G$ and $\lambda$:
\[
P^{\rm hyp}(z)=-G\alpha_G\lambda^2e^{-z/\lambda}
F(\rho_i^{(1)},\rho_j^{(2)},\lambda),
\]
\noindent
where the explicit form for $F$ is determined by the structure of
both plates which usually consist of several layers with the
densities $\rho_i^{(1)}$ and $\rho_j^{(2)}$. The Newtonian
gravitational force acting between closely spaced plates is
negligibly small.

The constraints on ($\lambda,\alpha_G$) following from the two
measurements of the Casimir pressure by means of micromechanical
torsional oscillator are shown in figure 14 by lines 1a and 1b. 
Line 1b is obtained from the first experiment \cite{40}, and
line 1a from the second, improved experiment \cite{41,52}.
For comparison, in the same figure the constraints following from
old Casimir force measurements between dielectrics \cite{7,9}
are shown by line 2, from torsion pendulum experiment \cite{11,53}
by line 3 (this is a continuation to smaller $\lambda$ of the line 7
in figure 13), and from the experiment by means of an atomic force
microscope \cite{17,54} by line 4. As is seen from figure 14, the
constraints obtained from the measurements of the Casimir force
by means of micromechanical torsional oscillator completely fill
in the gap between the modern constraints obtained using an atomic 
force microscope, and those obtained using a torsion pendulum.

Further work is needed to significantly strengthen the constraints
on the predictions of extra-dimensional physics and other extensions
of the standard model in a wider interaction range. For this purpose,
experiments over a wider separation range which use smoother and thicker 
metal coatings on the surfaces of the test bodies are planned.

\section{Conclusions}

As is evident from the foregoing, the Casimir effect is the subject
of diverse theoretical and experimental studies, and applications
in both fundamental physics and nanotechnology. This is due to the
fact that the concept of zero-point vacuum
fluctuations, central to the Casimir effect, is one of the most
general, fundamental, and even puzzling concepts of modern physics.
There are many other important applications of the Casimir effect
in elementary particle physics, gravitation and cosmology, atomic
and condensed matter physics which can not be discussed here due to 
the lack of space. They are connected with the bag model of hadrons,
mechanisms of spontaneous compactification of extra dimensions,
inflationary cosmology, problem of dark matter, theory of Rydberb
atoms, atomic friction, wetting phenomena etc. \cite{7,8,9,10}.

In our opinion, the most striking development of the last few years
on the subject is the precision measurements of the Casimir force
between metal surfaces. The novel experimental approaches have
opened up new fields of application of dispersion forces and
called for further theoretical investigations in the case of
real boundaries. The puzzle of the thermal Casimir force between real 
metals which was discovered at this point calls for a few additional
remarks. In fact, the most of theoretical output in the Casimir
effect before the year 1997, when the intensive experimental work
had begun, was produced by field theorists. The standard pattern of
research in this area is the use of exact mathematical tools
in the framework of physical theory applied to a rather simple
model of a system under study. It was desirable that the model used
provide full correspondence to all relevant properties of the 
physical system. This ideal pattern is, unfortunately, unattainable
in condensed matter physics. The physical systems under consideration
there (real metals, for instance) are so complicated that one must
consider a set of different approximate models to theoretically describe
different properties of a system. It is self-evident that in doing so 
all basic principles of quantum mechanics, quantum electrodynamics,
thermodynamics and statistical physics must be preserved. This
preservation, however, is not automatic. If the model used leads to
violation of some basic principle (as occurs in the Drude model
approach to the thermal Casimir force) this should be considered
as a signal that the model under consideration does not reflect 
some feature of the physical system important for the phenomenon
of our interest. At the same time, this model can be entirely
adequate in
certain physical situations (Drude model, for instance, provides
good description for many electrical properties of metals and
for optical processes with real photons). Another model
(the Leontovich impedance in our case) may be adequate in the area
where the Drude model fails. At present, several experiments to
measure the small thermal effect in the Casimir force have been
proposed \cite{55,56,57,58}. The performance of these
experiments will bring final resolution to the puzzle discussed
in this paper.

We expect that the role of the van der Waals and Casimir forces 
in nanotechnology will increase. Miniaturization is the
main tendency in modern technology and, thus, below some
transition size-scale, these forces will become dominant.
The application of the Casimir force as a test for fundamental
physics also seems to be very promising. The relatively cheap,
table-based laboratory experiments on the Casimir effect may be
considered as an alternative for huge accelerators.
All this suggests that we are only at the threshold of
an important scientific line of inquiry devoted to the
Casimir effect. 

\section*{Acknowledgements}

We are grateful to Ricardo Decca, Ephraim Fishbach, Dennis 
Krause and Umar Mohideen for numerous informative discussions.
We also acknowledge kind hospitality at the Department of Physics 
of the University of California (Riverside) where this paper
was written.


\begin{figure}
\vspace*{-7cm}
\centerline{
\includegraphics{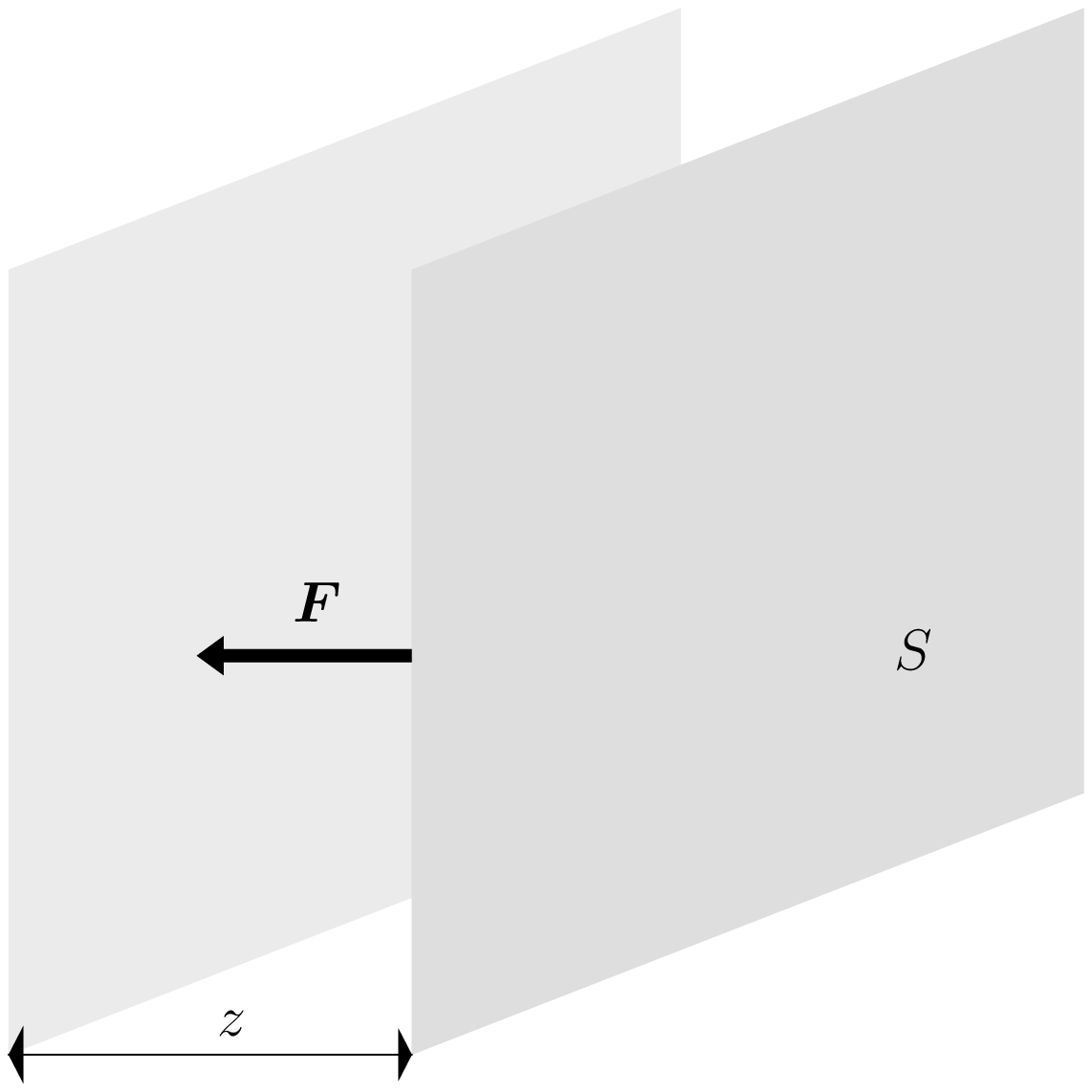}
}
\vspace*{-9cm}
\caption{Configuration of two parallel plates of area $S$ made
of ideal metal spaced $z$ apart.}
\end{figure}
\begin{figure}
\vspace*{-7cm}
\centerline{
\includegraphics{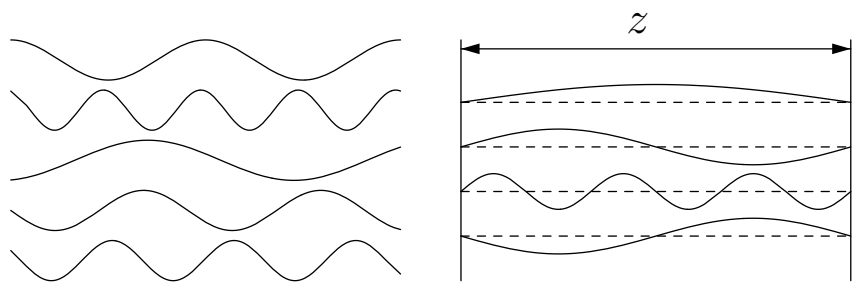}
}
\vspace*{-11cm}
\caption{Zero-point vacuum oscillations; left: in empty space;
right: in between two ideal metal plates.}
\end{figure}
\begin{figure}
\vspace*{-7cm}
\centerline{
\includegraphics{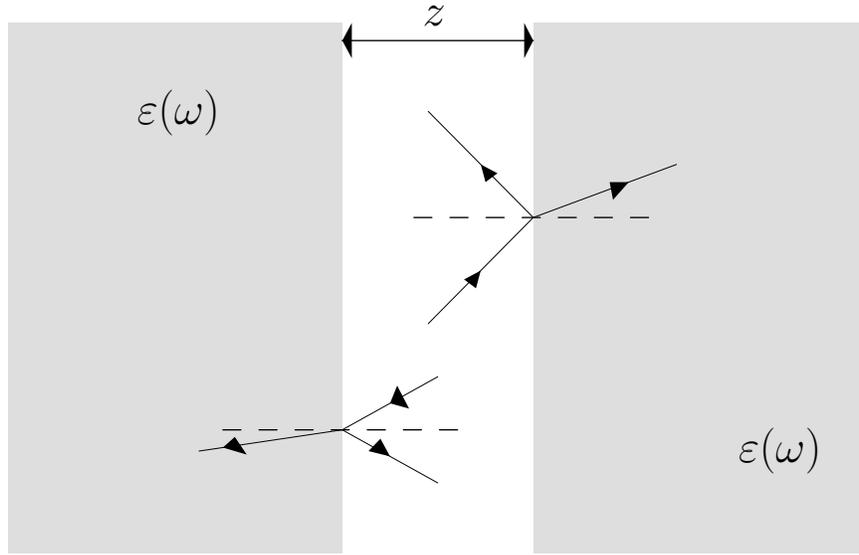}
}
\vspace*{-9cm}
\caption{Two semispaces made of some real material with the
dielectric permittivity $\varepsilon(\omega)$ and reflections
of electromagnetic oscillations on their boundary planes.}
\end{figure}
\begin{figure}
\vspace*{-4cm}
\centerline{
\includegraphics{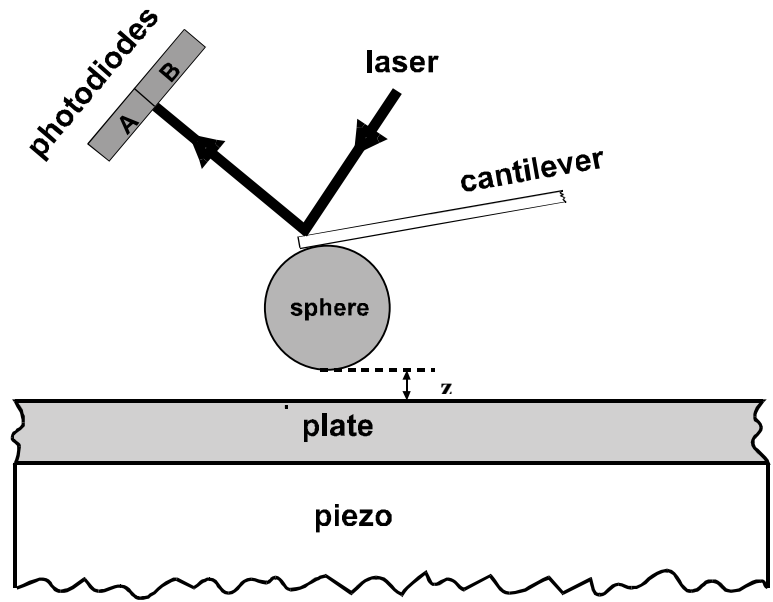}
}
\vspace*{-10cm}
\caption{Schematic diagram of the experimental setup for
measuring the Casimir force between a plate and a sphere using
an atomic force microscope.}
\end{figure}
\begin{figure}
\vspace*{-2cm}
\centerline{
\includegraphics{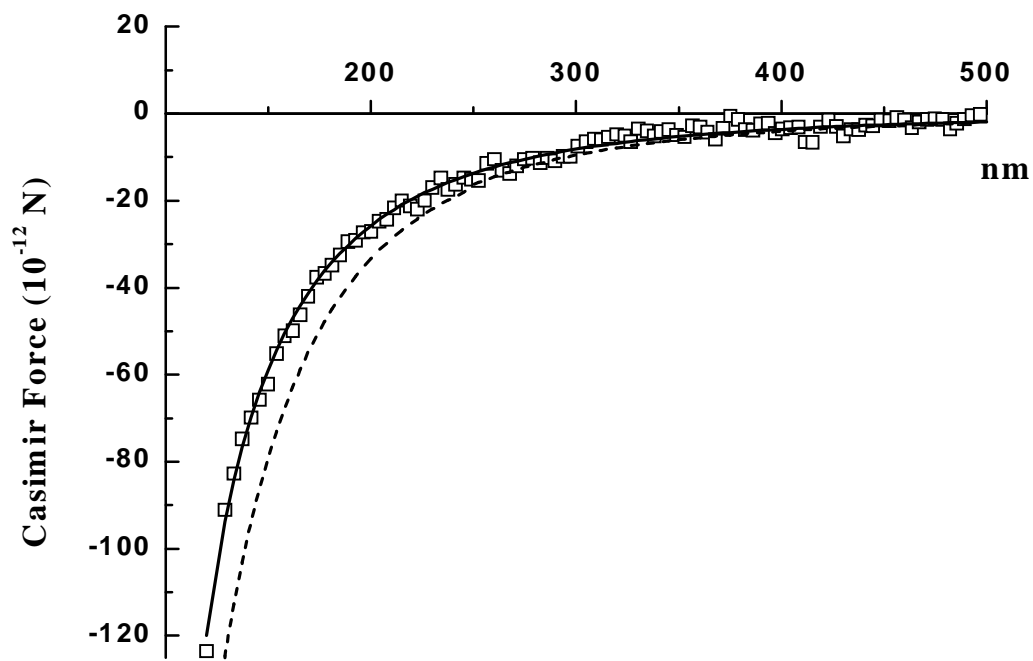}
}
\vspace*{-12cm}
\caption{Measured average Casimir force as a function of 
plate-sphere separation (open squares) compared with theory
taking into account corrections due to surface roughness and
finite conductivity (solid line) and without any correction
(dashed line).}
\end{figure}
\begin{figure}
\vspace*{-3cm}
\centerline{
\includegraphics{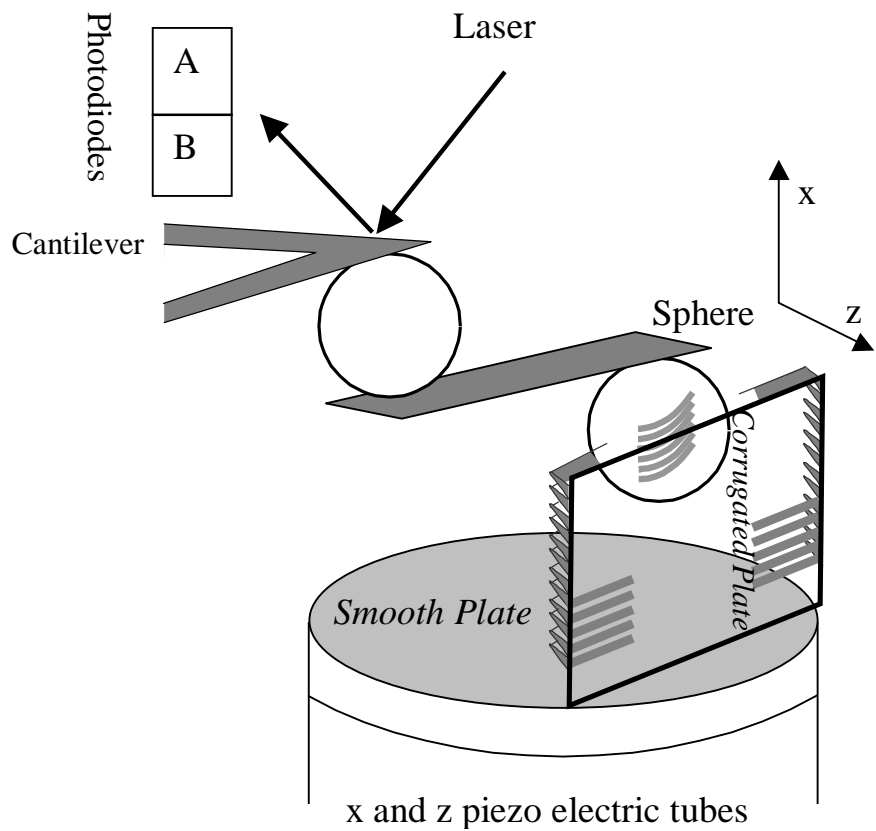}
}
\vspace*{-9cm}
\caption{Schematic diagram of the experimental setup for
measuring the lateral Casimir force between the corrugated plate
and the sphere using an atomic force microscope.}
\end{figure}
\begin{figure}
\vspace*{-4cm}
\centerline{
\includegraphics{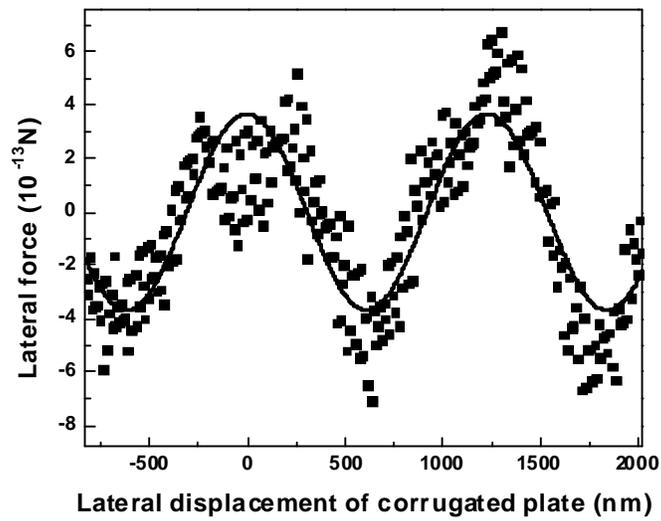}
}
\vspace*{-9cm}
\caption{Measured average lateral Casimir force as a function of
the lateral displacement between corrugations on a sphere and
a plate (solid squares) compared with theory (solid line).}
\end{figure}
\begin{figure}
\vspace*{-5cm}
\centerline{
\includegraphics{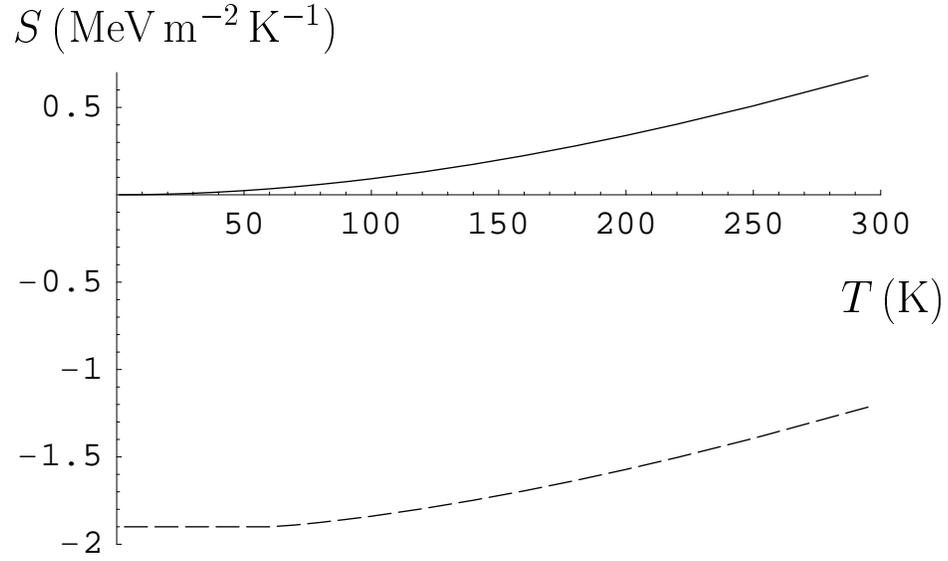}
}
\vspace*{-10cm}
\caption{Casimir entropy per unit area
as a function of temperature at a separation 1\,$\mu$m
between plates 
computed using the plasma model (solid line) and the
Drude model (dashed line).}
\end{figure}
\begin{figure}
\vspace*{-1cm}
\centerline{
\includegraphics{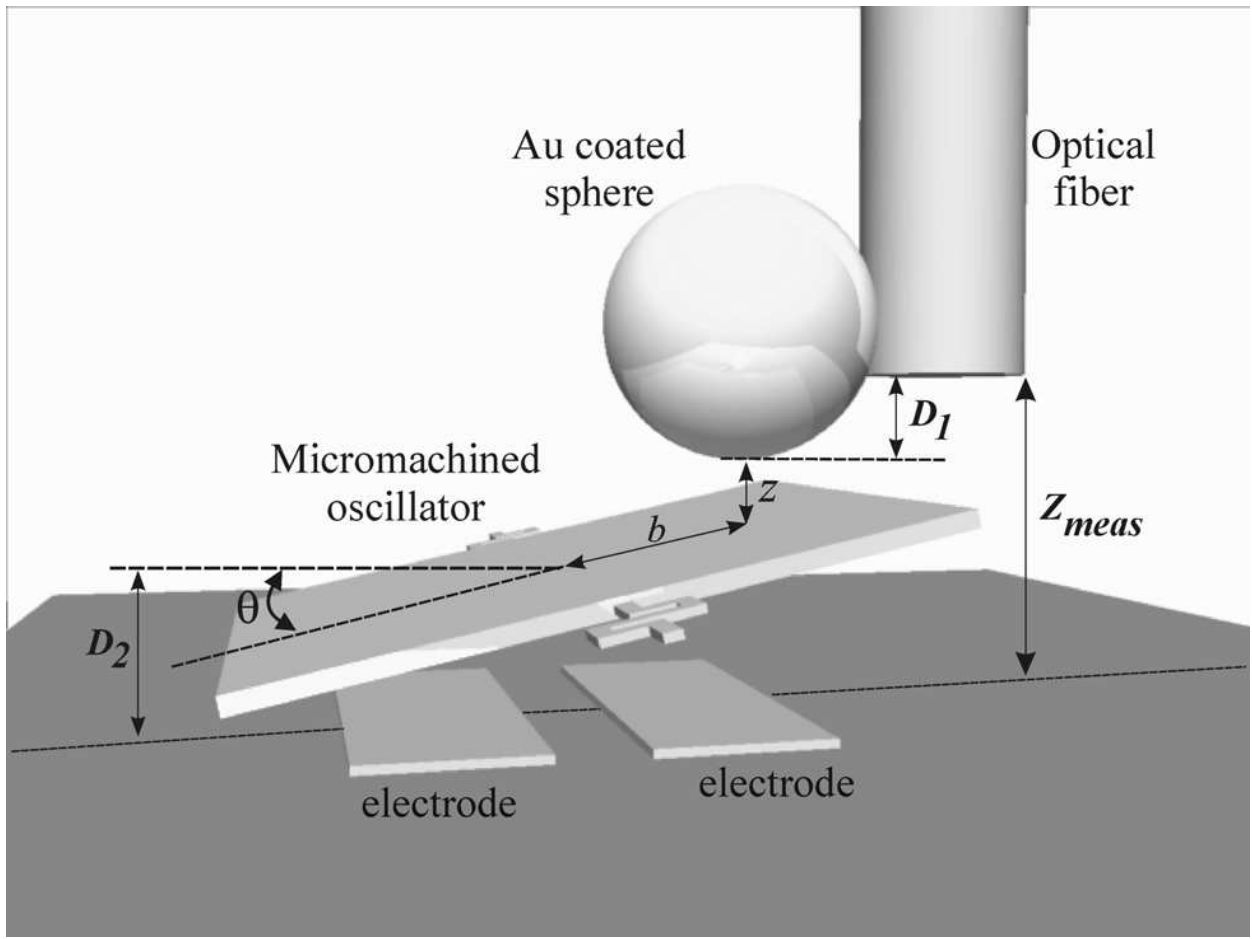}
}
\vspace*{-10cm}
\caption{Schematic diagram of the experimental setup for the
determination of the Casimir pressure between two parallel gold
coated plates using a micromechanical torsional oscillator.}
\end{figure}
\begin{figure}
\vspace*{-4cm}
\centerline{
\includegraphics{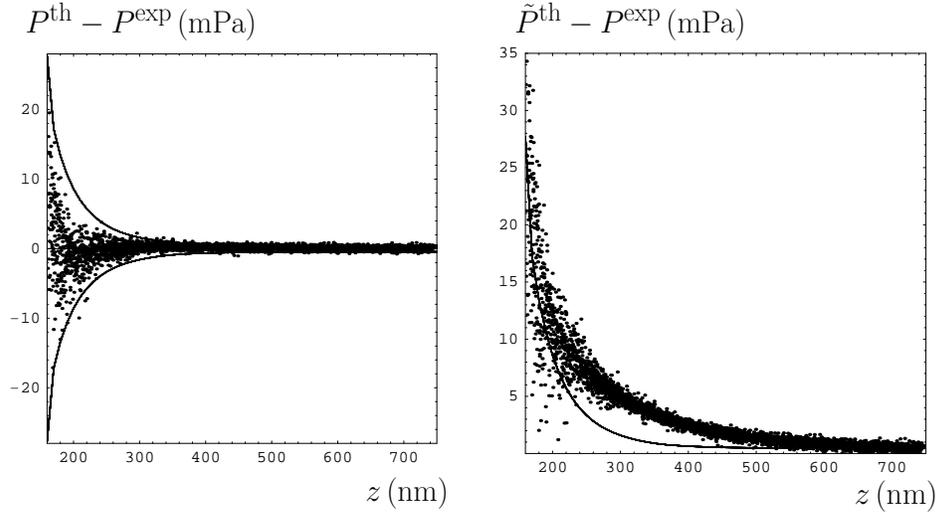}
}
\vspace*{-12cm}
\caption{Differences between theoretical and experimental Casimir
pressures between gold coated plates (dots) and the absolute errors
of these differences at 95\% confidence 
(solid lines) as functions of separation;
left: theory uses the impedance approach; right: theory uses the
Drude model approach.}
\end{figure}
\begin{figure}
\vspace*{-7cm}
\centerline{
\includegraphics{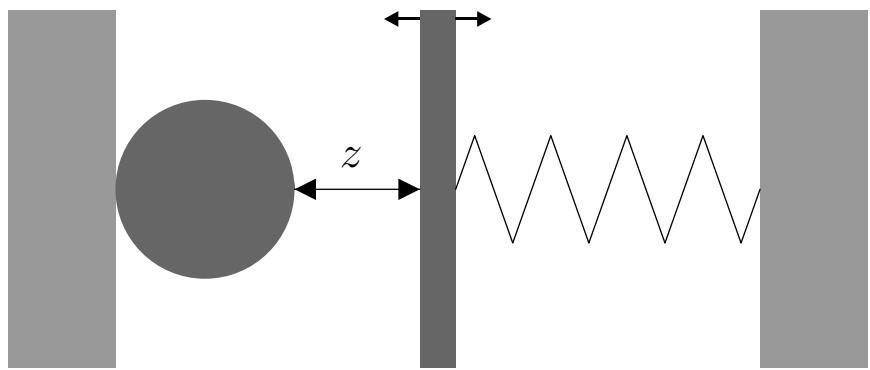}
}
\vspace*{-9cm}
\caption{Schematic diagram for the model of oscillator actuated
by the Casimir force.}
\end{figure}
\begin{figure}
\vspace*{-6cm}
\centerline{
\includegraphics{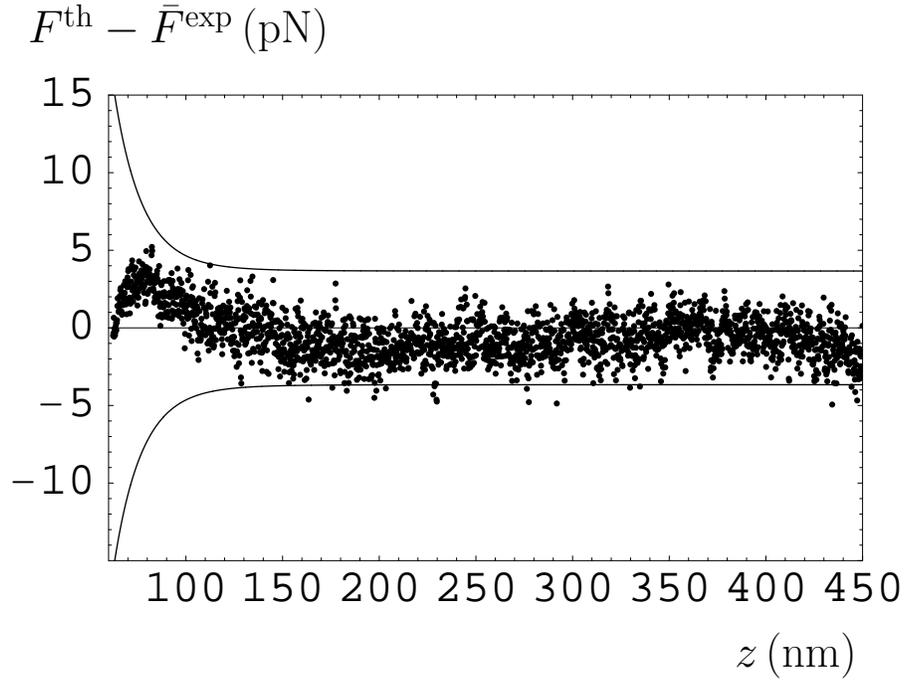}
}
\vspace*{-9cm}
\caption{Differences of theoretical and mean measured Casimir
force acting between silicon plate and gold coated sphere (dots) 
and the absolute errors of these differences at 95\% confidence 
(solid lines) as functions of separations.}
\end{figure}
\begin{figure}
\vspace*{-6cm}
\centerline{
\includegraphics{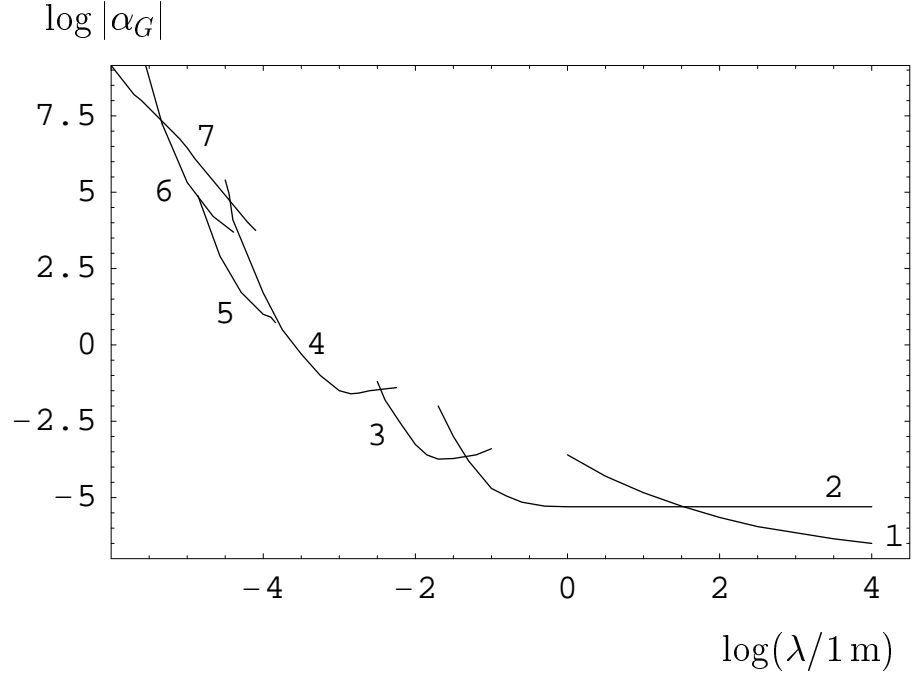}
}
\vspace*{-9cm}
\caption{Constraints on the strength of the Yukawa interaction as
a function of interaction range obtained from the experiments
of E\"{o}tvos-type (lines 1,\,2), Cavendish-type (lines 3--6),
and from the measurement of the Casimir force by means of
torsion pendulum (line 7).}
\end{figure}
\begin{figure}
\vspace*{-6cm}
\centerline{
\includegraphics{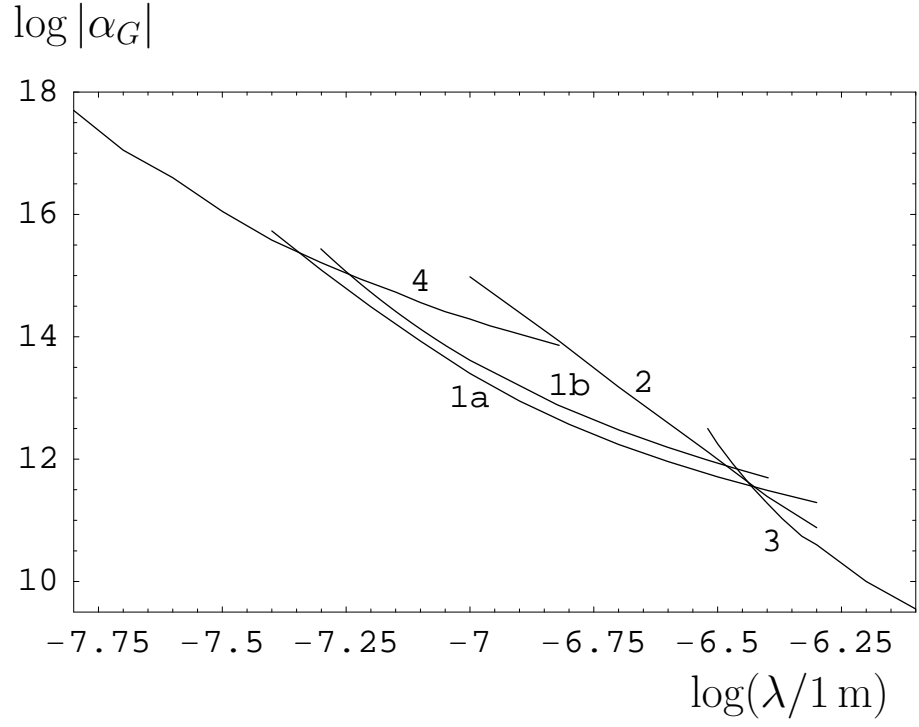}
}
\vspace*{-8cm}
\caption{Constraints on the strength of the Yukawa interaction 
obtained from the measurements of the Casimir force by means of
micromechanical torsional oscillator (lines 1a and 1b) and
earlier experiments (lines 2--4). See text for a more detailed 
characterization of the lines.}
\end{figure}
\end{document}